\documentclass[aps,twocolumn,showpacs,superscriptaddress,groupedaddress]{revtex4}  
\usepackage{graphicx}  
\usepackage{dcolumn}   
\usepackage{bm}        
\usepackage{amssymb, amsmath, amsfonts}   
\usepackage{lipsum}        
\usepackage{color}

\DeclareGraphicsExtensions{{.pdf}}
\graphicspath{{./}{./figure}}
\def\Re{\sf Re}
\def\bF{\boldsymbol{F}}

\begin{document}
\newcommand{\note}[1]{{\color{red}{#1}}}
\newcommand{\noter}[1]{{\color{red}{#1}}}
\newcommand{\noteb}[1]{{\color{blue}{#1}}}

\widetext


\title{
Hydrodynamic synchronization of
externally driven colloids
}
\author{Norihiro Oyama}
\affiliation{Mathematics for Advanced Materials-OIL, AIST-Tohoku University, Sendai 980-8577, Japan}
\affiliation{Department of Chemical Engineering, Kyoto University, Kyoto 615-8510, Japan}
\author{Kosuke Teshigawara}
\author{\\John Jairo Molina}
\affiliation{Department of Chemical Engineering, Kyoto University, Kyoto 615-8510, Japan}
\author{Ryoichi Yamamoto}
\affiliation{Department of Chemical Engineering, Kyoto University, Kyoto 615-8510, Japan}
\affiliation{Institute of Industrial Science, The University of Tokyo, Tokyo 153-8505, Japan.}
\author{Takashi Taniguchi}
\affiliation{Department of Chemical Engineering, Kyoto University, Kyoto 615-8510, Japan}
\date{\today}

\begin{abstract}
    The collective dynamics of externally driven $N_{\rm p}$-colloidal systems ($1\le N_{\rm p}\le 4$) in a confined viscous fluid have been investigated using three-dimensional direct numerical simulations with fully resolved hydrodynamics.
    The dynamical modes of collective particle motion are studied by changing the particle Reynolds number {as determined by the strength of the external driving force} and the confining wall distance.
    For a system with $N_{\rm p}= 3$,
    we found that at a critical Reynolds number, a dynamical mode transition occurs from the doublet-singlet mode to the triplet mode, which has not been reported experimentally.
    The dynamical mode transition was analyzed in detail
    from the following two viewpoints: (1) spectrum analysis of the time evolution of a tagged particle velocity and (2) the relative acceleration of the doublet cluster with respect to the singlet particle.
    For a system with $N_{\rm p}=4$,
    we found similar dynamical mode transitions from the doublet-singlet-singlet mode to the triplet-singlet mode and further to the quartet mode.
\end{abstract}

\maketitle

\setcounter{section}{0}
\section{Introduction}
Dispersions of solid colloidal particles can be found 
in various situations both in nature and in engineering applications, 
such as muddy flows, fluidized beds and sedimentation processes\cite{Russel1989a,Fan1998a}.
When particles are dispersed in a fluid, they interact with each other not only through direct forces, {\it e.g.}, the Van der Waals and electrostatic forces, but also indirectly through the surrounding fluid.
Such an indirect interaction through the fluid is known as hydrodynamic
interaction\cite{Russel1989a},
which makes the dynamics of particles complicated
due to its long-ranged nature and nonlinearity.
Although the inertia effect can be neglected, making the governing equation for the fluid flow linear when the Reynolds number ($\Re$) is very low (referred to as the Stokes regime),
it is known that even at such very small Reynolds numbers,
the long-ranged nature can lead to nontrivial collective particle behaviors.
For example, Reichert and Stark showed theoretically that a nontrivial limit cycle motion of three particles can appear as a stable steady state due only to hydrodynamic interaction in the limit of ${\Re} \to 0$\cite{Reichert2004}.
Of special importance here is that the authors considered a system with a very simple geometry of a quasi-one-dimensional trajectory in which the particles are bound to a circular path and driven along the path by a constant external force.
It is quite striking that even in such a simple system and under the linear condition,
the long-ranged nature of the hydrodynamics can lead to a nontrivial collective motion of particles.

Recently, great developments in manipulation techniques for microscopic systems, such as the optical trap/tweezers\cite{Curtis2003,Huang2011} and microfluidics\cite{Beatus2006,Beatus2007,Ito2017}, have enabled us to conduct well-controlled experiments at the micrometer scale.
In particular, optical tweezers have been applied to colloidal dispersions rather frequently\cite{Jabbari-Farouji2007,Yang2009,Curtis2003,Lutz2006,Roichman2007,Sokolov2011,Sassa2012}.
The work by Reichert and Stark\cite{Reichert2004} is motivated by the experiment by Curtis and Grier\cite{Curtis2003}, where a vortex optical tweezer was employed to drive the colloidal particles along a circular path.
In experiments on a three-particle system, a limit cycle motion is observed.
First, two of three particles moving on the circular path form a doublet cluster, {the velocity of which is faster than that of the remaining single particle due to hydrodynamic interactions.
The cluster is then able to catch up with the single particle, and the particles form a transient three-particle cluster triplet.}
Then, the front two particles in this triplet cluster detach, leaving the rear particle behind, and the system again consists of a new doublet cluster and a single particle.
The system repeats this cyclical motion.
Sassa {\it et al.} investigated a similar system but with different number of particles and observed similar collective motions depending on the number of particles\cite{Sassa2012}. 
Though these previous works successfully revealed the existence of 
the hydrodynamic coupling of dynamics of particles in the Stokes regime, 
the effects of the strength of the driving force have not yet been studied in depth.
The current techniques do not allow us to conduct well-controlled experiments over a wide range of Reynolds numbers.
Therefore, we still do not know what occurs when the Reynolds number becomes large and the system is in the nonlinear regime, where the theoretical approach using the mobility tensor is not applicable even for the very simple system mentioned above.
{In addition, for the system where confinements are introduced,
the explicit general form of the mobility tensor is still unknown, and
the behavior of the system is not theoretically predictable.}

In the present work, we
numerically study similar but simpler systems than
those in the previous works\cite{Sassa2012,Reichert2004}.
We consider the dynamics of particles driven along a straight
line under periodic boundary conditions.
The dynamics are quasi one-dimensional, and
particles do not rotate\cite{note:rotation}.
The number of particles $N_{\rm p}$ considered here is in the range $1\le N_{\rm p}\le 4$.
For such systems, we perform three-dimensional (3D) direct
numerical simulations with full hydrodynamics and investigate the dynamics of particles by changing the strength of
the external driving force.
Furthermore, we investigate the confinement effect on the dynamics of particles by introducing flat parallel walls explicitly and varying the separation
of the walls.
Although a previous work\cite{Roichman2007} has reported that particle dynamics are affected by the spatial inhomogeneity in the intensity of the optical vortex,
in the present work, we assume that the homogeneity of the intensity is as in the literature \cite{Reichert2004} to study the purely hydrodynamic effect.
We vary the external force widely so that the system
can reach the Allen regime,
where we can see a nonlinear Reynolds number dependence of the drag coefficient of an isolated single spherical particle.
As a result, it is clarified that the size of the stable cluster changes depending on the Reynolds number.
Furthermore, we observe the dynamical mode transition in all cases.
This transition is investigated in detail by conducting a spectrum analysis of the time evolution of the velocity of a particle, and it is understood in analogy with the second order phase transition.
Moreover, the dynamical mode transition is also studied from the viewpoint of the stability of the cluster by numerically evaluating the total force on each particle.
\begin{figure}
  \begin{center}
\includegraphics[width=\linewidth]{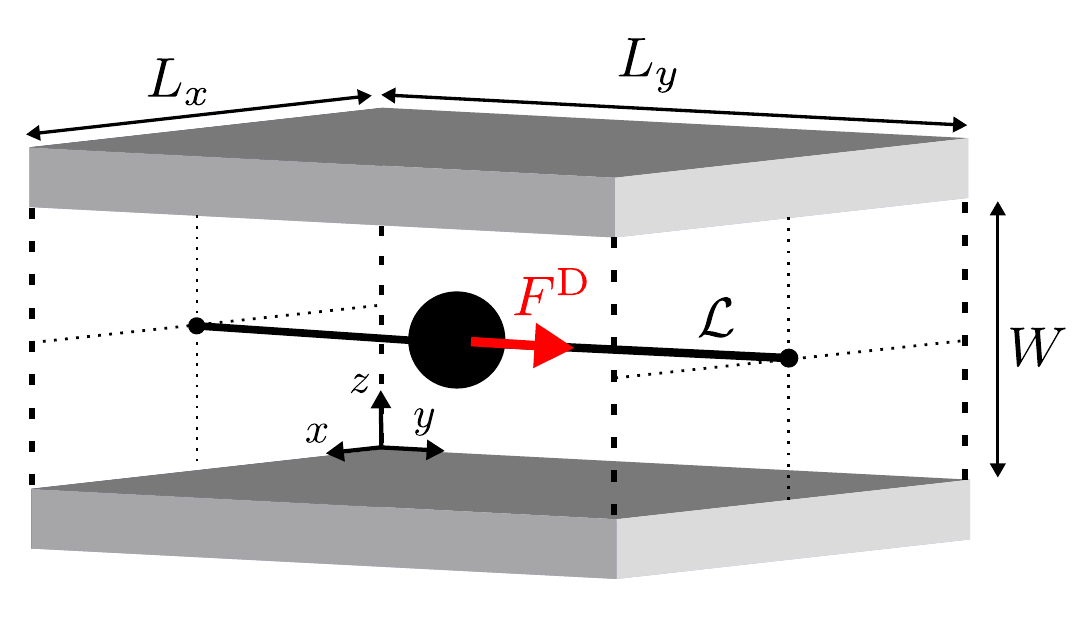}
\caption{\label{fig:system}
Schematic of the system. 
The straight line ${\cal L}$ stands for the path along which the
particle is driven.
}
\end{center}
\end{figure}

\section{Modeling \& Methods}\label{sec:model}
We consider externally driven spherical particles immersed
in a Newtonian fluid with a constant viscosity.
Although the particles are driven along a circular path
in the previous experiments\cite{Sassa2012,Reichert2004},
in the present work, we consider a more simplified path, {\it i.e.}, a straight path.
As shown in Fig.\ref{fig:system}, 
the system is confined by two flat parallel
walls that are placed at $z=0$ and $W$ and are perpendicular to the $z$-axis.
In the $x$ and $y$ directions, the system is periodic
(except for the cases shown in Sec.~\ref{sec:p1_p2},
where we impose periodic boundary conditions
in all directions with cubic computational domain).
The linear dimensions of the systems are
$L_x$, $L_y$ and $L_z$ in each direction
(when walls are introduced, $L_z\equiv W$).
We put particles
on the center line ${\cal L}$, as shown in Fig. 1,
and investigate the hydrodynamically mediated collective motion of the particles
under an externally applied constant driving force.
Throughout the present paper,
we ignore thermal fluctuations ({\it i.e.}, this corresponds to the limit of an infinite Peclet number\cite{note:Peclet}). 
The particles are identical to each other and are spherical with a diameter $D$.
In the simulations, particles are driven by a constant force along the
straight line ${\cal L}$.
Because the centers of mass of the particles are all
placed on the line $\cal L$ initially and no thermal noise is applied,
the particle trajectories will never deviate
from the line ${\cal L}$.
The diameters of particles are fixed to
$D=6\Delta$, where $\Delta$ stands for
the grid spacing and is the unit of length in our calculation.
The linear dimensions of the system in the $x$ and $y$ directions are set to be
large enough to ignore the influence of the periodic
boundary condition, $L_x=L_y=128\Delta = \frac{64}{3}D$.
We varied the distance between the two parallel walls, $W$, in the range of $2D\le W \le \frac{62}{3}D$; {\it i.e.},
ranging from a gap comparable to the particle diameter to that comparable to
the linear dimension in the $x$ and $y$ directions.
The driving external force $\bF^{\rm D}$ along ${\cal L}$ is expressed as:
\begin{align}
  \bF_{i}^{\rm D}
  = F^{\rm D}\hat{\boldsymbol{e}}_{\rm y},
\end{align}
where
$i$ is the particle index and 
$\hat{\boldsymbol{e}}_{\rm y}$ the unit vector in the $y$ direction.
Note that $F^{\rm D}$ is the same for all particles.
The driving force $F^{\rm D}$ used in this work leads to
$\Re\le 10$ for all the $W$ values used.

To simulate the dynamics of the system numerically,
one must solve the equations of motion of the particles and the hydrodynamic equation for the host fluid simultaneously.
The dynamics of the particles are described by the Newton-Euler
equations of motion:
\begin{align}
  \dot{\boldsymbol{R}}_i &= \boldsymbol{V}_i ,\label{eq:N-E}\\
  M_{\text{p}}\dot{\boldsymbol{V}}_i &=
  \boldsymbol{F}_i^{\text{H}}+\boldsymbol{F}_i^{\text{P}}+\boldsymbol{F}^{\rm D}_i,\\
  \dot{\boldsymbol{Q}}_i &= \text{skew}(\boldsymbol{\Omega}_i)\cdot
  \boldsymbol{Q}_i,\\
  \boldsymbol{I}_{\text{p}}\cdot\dot{\boldsymbol{\Omega}}_i &= \boldsymbol{N}_i^{\text{H}}
  , \end{align}
where  $\boldsymbol{R}_i$ is
the position, $\boldsymbol{V}_i$ the velocity,
$\boldsymbol{\Omega}_i$ the angular velocity and
$\boldsymbol{Q}_i$ the
rotation matrix of particle $i$.
The mass and the moment of inertia are expressed as $M_{\text{p}}$
and $\boldsymbol{I}_{\text{p}}$, respectively,
and $\text{skew}(\boldsymbol{\Omega}_i)$ is the skew-symmetric
angular velocity matrix.
The particle motion is coupled with the fluid flow through 
the hydrodynamic force $\boldsymbol{F}_i^{\text{H}}$
and torque $\boldsymbol{N}_i^{\text{H}}$
exerted by the ambient fluid on the particle $i$.
To avoid a particle overlap, which can occur
owing to the finite temporal and spatial discretization,
a direct particle-particle interaction,
$\boldsymbol{F}_i^{\text{P}} $, is introduced by using a truncated
Lennard-Jones potential with a power of 36 for the repulsive part,  as follows:
\begin{align}
  \boldsymbol{F}_i^{\rm P} &= \sum_j\boldsymbol{F}^{\rm P}_{ij},\quad
  \boldsymbol{F}^{\rm P}_{ij} = -\boldsymbol{\nabla}_i U\left(r_{ij} \right),\label{Eq:Fc}\\
  U\left( r_{ij}\right) &=
  \begin{cases}
    4\epsilon\left[ \left( \frac{D}{r_{ij}}\right)^{36} -
      \left( \frac{D}{r_{ij}}\right)^{18}\right] + \epsilon
    &\left( r_{ij}\le  r_{\rm C}\right) \\
    0 & \left( r_{ij}\ge r_{\rm C}\right)
    \end{cases}
\end{align}
where $\boldsymbol{F}^{\rm P}_{ij}$ stands for the direct particle-particle interaction force exerted on particle $i$ by particle $j$, $r_{ij}$ the distance between the two particles, $r_{\rm C}$  represents the cutoff length, and
$\epsilon$ represents the strength of the potential and the energy unit of the system.
Because the potential changes steeply within the cutoff length $r_{\rm C}$,
the value of $\epsilon$ does not affect the whole dynamics of the system.
As the particle trajectories never deviate from the center line ${\cal L}$, as mentioned above, and they never approach the walls, there is no need to take into account the excluded volume effect between the particles and walls.
The fluid flow is described by the Navier-Stokes equation;
\begin{align}
  \rho_{\text{f}} \left( \partial_t + \boldsymbol{u}_{\rm
    f}\cdot\nabla\right)\boldsymbol{u}_{\rm f}
                &= \nabla \cdot \boldsymbol{\sigma}_{\rm f},\\
    \boldsymbol{\sigma}_{\rm f} &=
    -p\boldsymbol{I}+\eta\left\{
    \boldsymbol{\nabla}\boldsymbol{u}_{\rm f}
    +\left( \boldsymbol{\nabla}\boldsymbol{u}_{\rm f}\right) ^t
    \right\}\label{eq:N-S},
\end{align}
with the
incompressible condition:
\begin{align}
  \boldsymbol{\nabla}\cdot\boldsymbol{u}_{\rm f} &= 0,  
\end{align}
where $\boldsymbol{u}_{\rm f}$ is the fluid velocity field,
$\boldsymbol{\sigma}_{\rm f}$ is the Newtonian stress tensor,
$p$ is the isotropic pressure, $\boldsymbol{I}$ is the unit tensor,
$\rho_\text{f}$ is the mass density and $\eta$ is the shear viscosity of the host fluid.
To couple the equations (\ref{eq:N-E})-(\ref{eq:N-S}),
one has to handle moving
sharp boundaries between solid particles and the host fluid.
It is known that the treatment of such sharp boundaries is
computationally expensive.
To address the boundaries more
efficiently, we employ the Smoothed Particle Method (SPM)\cite{Nakayama2005,Kim2006,Nakayama2008,Molina2013b}, which is
a well-established method for solid-fluid mixtures
(for this purpose, many other options are available:  fluid-particle dynamics\cite{Tanaka2000}, the lattice Boltzmann method\cite{Ladd1993,Alarcon2013},
the distributed Lagrange multiplier/fictitious domain method\cite{Glowinski1998,Glowinski2000},
multi-particle collision dynamics\cite{Allahyarov2002,Blaschke2016a},
and the boundary element method\cite{Youngren1975,Zhu2013}
).
In the SPM, such a sharp boundary between the phases is not considered
explicitly, but instead, a diffuse interface is introduced.
The solid and fluid phases
are distinguished by a phase field, $\phi$, that
varies across the interface region with a finite width of
$\xi$~(in this work, we use $\xi=2\Delta$ for all calculations).
The phase field takes $\phi=1$ in the solid domain and
$\phi=0$ in the fluid domain.
The walls are also expressed by $\phi=1$ in the domain inside the wall.
Using the phase field $\phi$, 
the total velocity field $\boldsymbol{u}$ is expressed as $\boldsymbol{u} = \left(
1-\phi\right)\boldsymbol{u}_{\rm f}+\phi \boldsymbol{u}_p$, where
$\phi \boldsymbol{u}_{\text{p}} = \sum_{i}\phi_i
[\boldsymbol{V}_i+\boldsymbol{\Omega}_i \times
  (\boldsymbol{x}-\boldsymbol{R}_i)]$
is the contribution from the particle velocity field,
$\boldsymbol{x}$ is the position of interest
and $\left(
1-\phi\right)\boldsymbol{u}_{\rm f}$ is the contribution from the fluid
motion.
As the governing equations for the total fluid $\boldsymbol{u}$,
we employ a modified incompressible Navier-Stokes
equation:
\begin{align}
    \rho_{\text{f}} \left( \partial_t + \boldsymbol{u}\cdot\nabla\right)\boldsymbol{u}
                &= \nabla \cdot \boldsymbol{\sigma}_{\rm f} + \rho_{\text{f}} \phi\boldsymbol{f}_{\text{p}},\\
     \boldsymbol{\nabla}\cdot\boldsymbol{u} &= 0,
   \end{align}
where $\rho_{\rm f} \phi \boldsymbol{f}_{\rm p}$ is the body force generated from the
rigidity condition of particles.
Using this scheme, the dynamics of the system are calculated
efficiently with fully resolved hydrodynamics.

To summarize, using the methods presented above,
we investigated the collective
dynamics of externally driven colloids in a confined viscous fluid by changing the distance of walls $W$,
the external driving force $F^{\rm D}$ and the number of particles $N_{\rm p}$.
We considered $2D\le W \le \frac{62}{3}D$ and $F^{\rm D}$ in the
range, which leads to $\Re \le 10$.
Regarding the number of particles $N_{\rm p}$,
we considered the range of $1\le N_{\rm p} \le 4$.
Thermal fluctuations are ignored in all the simulations here,
assuming a large enough particle size to neglect the Brownian motion.
We define the Reynolds number {$\Re$} of the present system as
$\Re$$\left(v\right) = {\rho_{\rm f}} {vD}/\eta$, where $v$ is the maximum velocity of a tagged particle
{when the steady state is a limit cycle motion}.
Hereafter, force, velocity and time are
scaled by characteristic values, $F^{\rm D}_0$, $v_0$ and $t_0$, respectively.
The unit velocity $v_0$ is defined by $v_0=\eta/\rho_{\rm f}D$.
Therefore,
using the unit velocity, the Reynolds
number can also be expressed as $\Re$$\left( v\right)=v/v_0$.
The time unit $t_0$ is defined as $t_0 = D/v_0$.
The unit force $F^{\rm D}_0$ is set to be $F_0^{\rm D}=3\pi \eta D v^{*}$, where $v^{*}$ is given by ${\Re}\left( v^{*}\right)=\rho_{\rm f}v^{*} D/\eta=0.1$.

\section{Results \& Discussions}\label{sec:results}
\subsection{Single- and two-particle systems}\label{sec:p1_p2}
\begin{figure}
  \begin{center}
\includegraphics[width=\linewidth]{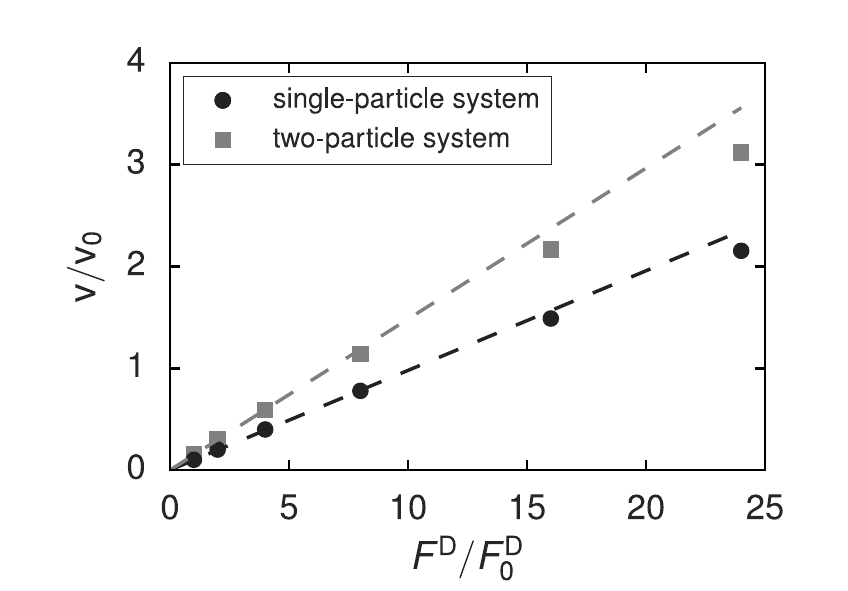}
\caption{\label{fig:p1_p2_vel}
  Steady state velocity of
  the singlet in a single-particle system and that of the 
  doublet in a two-particle system as functions of $F^{\rm D}$.
  Symbols stand for the simulation results.
  Lines are drawn to show the linearity of the data.
  The slopes of the lines are determined by the least squares method with the data below $\Re\lesssim1$
 (the four data points in the range of  $F^{\rm D}/F^{\rm D}_0<10$ for the single-particle system
    and the three data points in $F^{\rm D}/F^{\rm D}_0<5$ for the two-particle system).
}
  \end{center}
\end{figure}

In this section,
we demonstrate how hydrodynamic interactions can lead to
the collective motion of particles.
In the preceding experimental work~\cite{Sassa2012}
using a circular path,
it is reported that in a two-particle system,
one of the two particles catches up with the other spontaneously due to hydrodynamic interactions,
and the two particles move together as a stable cluster.
This dynamics is due to the fact that the rear particle feels less
hydrodynamic drag than the one in the front owing to the screening effect 
(although the path is circular, the front and the rear are defined by considering a pair with the shorter distance between particles).
Because of the difference in the strength of the hydrodynamic drag force,
the rear particle obtains a positive acceleration relative to the front particle,
and the rear particle will approach the front one, resulting in a cluster formation.
We refer to the cluster composed of two particles as a ``doublet'' and to the isolated particle as a ``singlet''. 
We performed simulations to verify whether a similar
cluster formation can be observed in our numerical systems using not a circular path but a straight path.
Note that in order to study the effect solely from the perspective of hydrodynamic interactions,
we set the periodic boundary conditions in all
directions and did not introduce any confinements for the simulations in this section,
though we introduce the confining walls in all other simulations in this work.
The linear dimensions of the system are $L_x=L_y=L_z=128\Delta\approx 20 D$.
As a result, we observed a doublet formation in
our system, in agreement with previous experiments..
We also confirmed that the doublet formation occurs irrespective of the initial configuration of the particles.
It is also reported in Ref.\cite{Sassa2012} that the velocity of the doublet is larger
than that of the singlet when $F^{\rm D}$ is the same.
To investigate the difference in the steady state velocity,
simulations for single- and two-particle systems are performed with
various $F^{\rm D}$ values.
In Fig.~\ref{fig:p1_p2_vel},
the steady state velocities of a singlet in the single-particle system and
that of a doublet in the two-particle system are plotted as functions of
$F^{\rm D}$.
The lines are drawn to visualize the linearity, which holds for small $\Re$.
The slopes are determined by the least squares method with the data below $\Re<1$
where the velocity is expected to be a linear function of $F^{\rm D}$.
As we can see in Fig.~\ref{fig:p1_p2_vel}, 
a doublet shows a velocity larger than a singlet for a given
$F^{\rm D}$, similar to the previous work\cite{Sassa2012}.
The ratios of the velocity of the doublet $v_2$ to that of the singlet $v_1$,
$v_2/v_1$ in the region $\Re < 1$ are approximately $v_2/v_1=1.5$,
which is close to the experimentally observed value, $v_2/v_1=1.3$.
Likewise, as will be seen in three- and four-particle systems later,
in the range of our consideration, $N_{\rm p}\le 4$,
the more particles a cluster contains,
the faster the velocity of the cluster becomes.

As shown above, particles driven along a periodic linear path form a
cluster through hydrodynamic interactions.
In addition, the steady state velocity of the doublet is larger than that of the singlet.
In the following sections,
the collective behaviors of three- and
four-particle systems will be presented
with the presence of parallel and flat confining walls.
We investigate the effects of the distance between the confining walls $W$
and the external force $F^{\rm D}$ on the
dynamical modes,
which has not been investigated by the preceding experiments.
In particular, we will report that new collective modes of particle dynamics emerge
in specific parameter regions.

\begin{figure*}
  \begin{center}
\includegraphics[width=\textwidth]{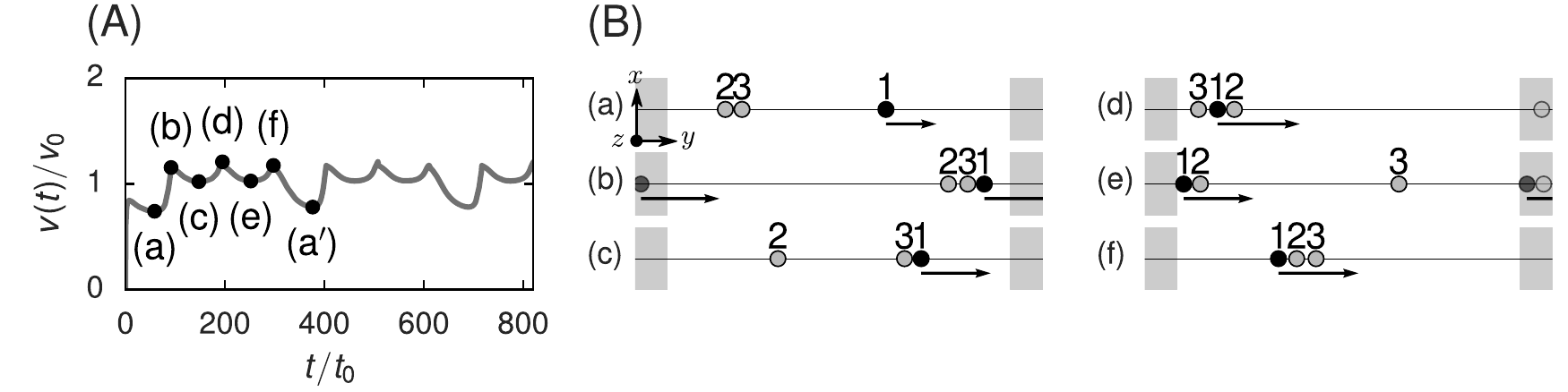}
\caption{\label{fig:sche_p3}
  Three-particle system with
  $W=\frac{62}{3} D$ and $F^{\rm
    D}=6F^{\rm D}_0$, where the doublet-singlet and triplet states are observed.
  (A) The time evolution of the velocity of a tagged particle.
  (B) Schematic snapshots of the system at
  (a)-(f) in (A).
  Here, pictures show only the 2D plane, which is parallel to the walls and
  includes the driving path ${\cal L}$.  
  The tagged particle is denoted by the black filled circle in (B).
  Shaded regions in (B)
  stand for copied simulation boxes according to 
  the periodic boundary condition, and the horizontal line at the center of
  each picture stands for the driving path, corresponding to the line
  ${\cal L}$ in Fig.~{\ref{fig:system}.
    In the driving direction (the $y$ direction),
    the full system is described, but in the other direction
  (the $x$ direction), only the vicinity of the driving path is depicted.
  }
  Arrows below the filled circles represent the velocity of the tagged particle, the length
  being proportional to the magnitude of the velocity.
}
  \end{center}
\end{figure*}

\subsection{Three-particle system}
We present the results for systems where three particles are towed along
a periodic straight line by a constant external force.
Here, to investigate the effect of confinement, two flat parallel walls with a separation distance $W$
are introduced.
In the initial configuration, the particles are set to be very close to
each other
(if particles are uniformly distributed initially at the exact same distances,
it will take a very long time for the system to reach a nontrivial collective motion).
We confirmed that systems with the same parameters but with different initial configurations fall into the same steady state.
In the previous experiments\cite{Sassa2012,Roichman2007,Lutz2006}, the existence of the
walls was not important because the distance between the
walls was much
larger than the particle diameter, $W\approx 30D$.
It is not obvious whether the confinement affects the
qualitative behaviors of the collective dynamics when the
distance between walls becomes comparable to the particle diameter.
Therefore, in this work, we investigated the collective motion of
particles for a wide range
of $W$, ranging from a distance comparable to the particle diameter to
a sufficiently larger distance similar to that in the previous experimental works.

  Before presenting the results of the current work, let us explain the experimentally observed collective motion of particles reported in the literature \cite{Sassa2012,Roichman2007,Lutz2006}.
  The authors reported a unique limit cycle collective motion for
  an externally driven system of three particles.
  The distance between the walls was fixed to be $W\approx 30D$.
  The driven path is a circular ring, the diameter $D_{\rm path}$ of which is set to be $\frac{10}{3}D\le D_{\rm
    path}\le 8D$.
  In such systems, first, a pair of particles with
  the smallest distance forms a doublet,
  which results in the particle configuration of one doublet and one
  singlet.
  Then, because the doublet is faster than the singlet,
  the doublet catches up to the singlet particle.
  Temporally, the particles form a triplet state in which all three particles move together as one cluster,
  but such a triplet state is unstable.
  The front two particles leave the rear particle and start to move as a doublet.
  Thus, the particle configuration is again composed of one doublet and
  one singlet.
  The system repeats this configurational change cyclically.

Now we would like to present the results of our simulations.
First, we investigated whether we can numerically reproduce
the experimentally observed collective motion mentioned in the above paragraph.
In Fig.~\ref{fig:sche_p3}, we show the time evolution of the velocity of a tagged
particle and corresponding schematic snapshots 
for the case with $W=124\Delta\approx 20D$ and $F^{\rm D}=6F^{\rm D}_0$.
The system with $W=124\Delta$ is the largest we have considered.
For this $W$,
the effect of the confinement can be assumed not to be important (this is quantitatively confirmed later in the discussion on Fig.~\ref{fig:p3_phase_diagram}),
and therefore, in this sense, this system corresponds to the previous experiment\cite{Sassa2012}.
As we can see from Fig.~\ref{fig:sche_p3}, a tagged particle shows a cyclic velocity change
with three velocity peaks, (b), (d) and (f), in every single cycle
(from (a) to ($\text{a}^\prime$), where (a) and ($\text{a}^\prime$) are identical
states with respect to the particle configuration).
Comparing the snapshots and the corresponding velocity
of the tagged particle,
we can see that a larger cluster has a larger velocity.
At the tops of these peaks, the system is in the triplet state, where
all three particles move together, nearly contacting with each
other.
We would like to note that unlike the doublet state
where two particles move together at the exact same velocity,
in this transient triplet state, all three particles do not always have the same velocity.
Therefore, we employed a definition of the triplet state 
from the view point of the configuration: we refer to the state as the ``triplet state''
when the maximum distance between any two adjacent particles is less than $1.1D$.
This threshold value can be set arbitrarily, but if it is set to be close to $1.1D$, the value itself does not affect the qualitative discussion below.
As we remarked, the triplet state ((b), (d) and (f)) is the fastest in the three
particle system.
However, such a state is unstable, and
the leading two particles start to move
as a doublet and leave
the remaining particle behind, which results in the ``doublet-singlet'' state
((a), (c) and (e)).
Because the system is periodic along the driven path,
the doublet catches up to the singlet particle,
and the system finally returns to the triplet state again.
The system repeats these configurational changes cyclically.
The behavior is qualitatively
the same as that reported in the experiments\cite{Sassa2012,Roichman2007,Lutz2006}.
We refer to this cyclic dynamical mode as the ``double-singlet mode.''
  
\begin{figure}
\includegraphics[width=\linewidth]{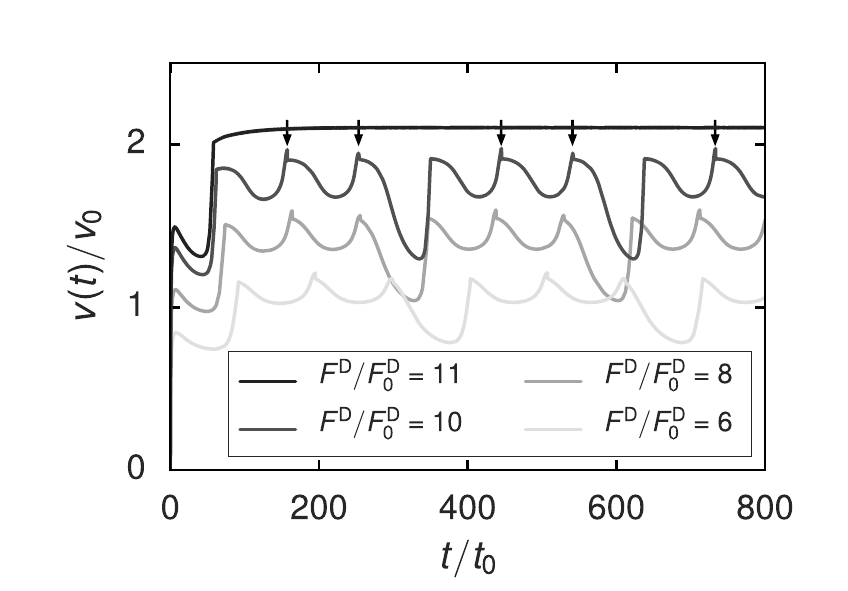}
\caption{\label{fig:p3_vel} Time evolution of the velocity of the tagged
  particle in the three-particle system with
  $W=\frac{62}{3}D$ for several $F^{\rm D}$s.
  Arrows indicate the positions of overshoots
  observed in the case with $F^{\rm
    D}=10F^{\rm D}_0$.
}
\end{figure}

So far, the discussion has focused on the case with the external force 
$F^{\rm D}=6F^{\rm D}_0$.
Next, we investigate the effect of the driving force by using various $F^{\rm D}$s
while keeping the distance between the two walls fixed at $W=\frac{62}{3}D$.
Similar to Fig.~\ref{fig:sche_p3}~(A), Fig.~\ref{fig:p3_vel} shows the
time evolutions of the velocity of a tagged particle 
for various $F^{\rm D}$s.
As the external force increases, the average velocity of the
tagged particle becomes larger, and
the triplet cluster has a longer life time 
~((b), (d) and (f) in Fig.~\ref{fig:sche_p3}),
as clearly seen in the case of $F^{\rm D}=10F^{\rm 0}_0$.
Furthermore, when the velocity exceeds a critical value,
the triplet state is stabilized completely, and
the system transits to a new dynamical mode where
the three particles move together stably with a constant velocity,
as shown in $F^{\rm D}=11F^{\rm D}_0$.
We refer to this dynamical mode as the ``triplet mode''.
The stabilization of the triplet state will be discussed quantitatively at the end of this section.
Note that we can see small and spiky velocity changes
at the second and the third peaks
(corresponding to the snapshots (d) and (f) in
Fig.~\ref{fig:sche_p3}(B))
for relatively high values of $F^{\rm D}$, as indicated by arrows for
the case with $F^{\rm D}=10F^{\rm D}_0$ in Fig.~\ref{fig:p3_vel}.
Such spikes can be explained by a two-step mechanism.
The doublet cluster containing the tagged particle has an increasing acceleration
when it is about to catch up to the leading particle
owing to the reduction in the hydrodynamic drag force.
Then, the doublet finally makes contact with the leading particle and starts moving
with the leading particle.
At the moment of contact, the velocity of the doublet jumps up and
then drops off due to the reaction force from the colliding singlet.
Such a spike is not observed {when the tagged particle is chosen to be the singlet particle that is approached by the doublet.}
  
\begin{figure}
\includegraphics[width=\linewidth]{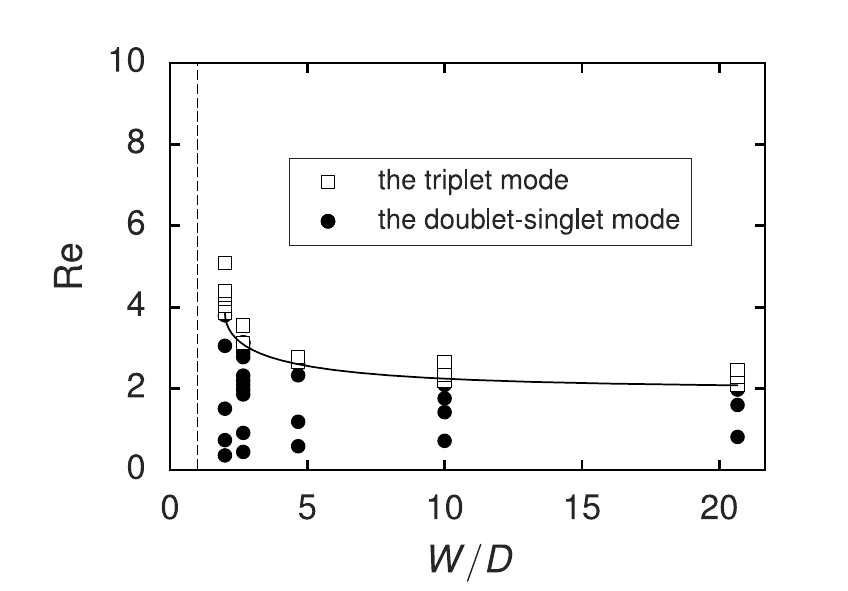}
\caption{\label{fig:p3_phase_diagram}
  Phase diagram of the dynamical mode for the three-particle systems.
  The filled circle stands for the doublet-singlet mode,
  and the open square represents the triplet mode.
  The solid line represents
  the transition line between 
  the two dynamical modes.
The dashed line denotes $W/D=1$, below which particles of size $D$ cannot be placed between the walls.}
\end{figure}

To investigate the wall separation dependence of the dynamical modes,
we also performed simulations for various values of $W$ and $F^{\rm D}$.
We used various $F^{\rm D}$s that lead to $\Re \le 10$.
The distance between walls ranges $2D \le W \le 62/3D$.
For any $W$ in this range, we observe two distinct dynamical modes, namely,
the triplet mode and the doublet-singlet mode.
In Fig.~\ref{fig:p3_phase_diagram}, we show the phase diagram of dynamical modes and 
the phase boundary line at which the dynamical mode transition occurs.
The Reynolds numbers used
in Fig.~\ref{fig:p3_phase_diagram} are calculated by using the maximum
velocity of a tagged particle.
This result means that the confinement does not change the qualitative features of
the collective motion in the three-particle system.
Quantitatively, 
we can see that the critical Reynolds number rises as $W$ decreases.
This is simply because of the suppression of the hydrodynamic
interaction by the stick boundary condition at the wall surfaces.
The narrower the wall separation becomes,
the stronger the hydrodynamic interaction is suppressed by the
walls.
Because of this suppression,
the hydrodynamic interaction between the particles
shows a faster spatial decay for smaller $W$ values.
Under a stronger confinement, a higher $\Re$ is needed to induce the transition from the doublet-singlet mode to the triplet mode.
  The detailed mechanism of the transition will be discussed later.

It has been well-investigated experimentally that
the dependence of the drag coefficient of a single spherical particle
on the Reynolds number
can be divided into three regimes:
the Stokes regime ($\Re \lesssim 2$), the Allen regime
($2 \lesssim \Re \lesssim 500$) and the Newton
regime ($500 \lesssim \Re$)\cite{Fan1998a}.
In the Stokes regime,
the governing equation of the system, the Stokes equation,
is linear with respect to the velocity
field and the motions of the particles can be handled
analytically by using the hydrodynamic force expressions given by
approximated mobility tensors, such as the Roton-Prager-Yamakawa tensor\cite{Rotne1969,Yamakawa1970,Wajnryb2013}.
In the Newton regime, the fluid flow becomes turbulent, and
the dynamics is chaotic. 
The Allen regime is referred to as the transient regime between
these two limiting
regimes because the dynamics are neither linear nor turbulent. 
Interestingly, the threshold Reynolds number between the Stokes
regime and the Allen regime matches the critical Reynolds number
between the two dynamical modes that we observe in the three-particle system,
$\Re \approx 2$, provided that the wall separation is large enough
($W \ge 10D$, where the wall effect can be ignored).
Although it is still an open question to clarify
the relation between the nonlinearity in the
Allen regime and the multi-particle
synchronicity in the three-particle system,
our results suggest the possibility that
these two phenomena share the same origin.

\begin{figure}
\includegraphics[width=\linewidth]{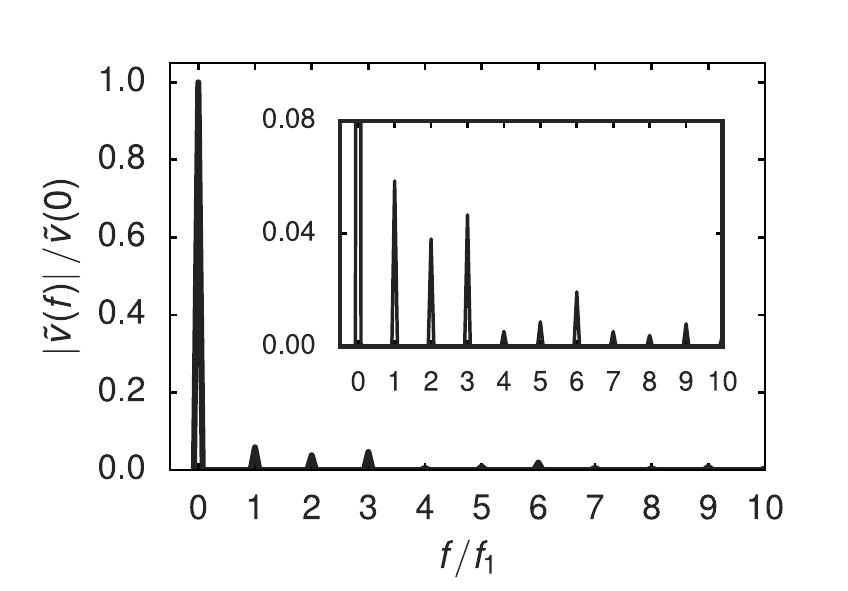}
\caption{\label{fig:fourier}
  Absolute values of Fourier coefficients of
  the velocity $v(t)$ of a tagged particle in the double-singlet mode of the three-particle
  system with
  $W=\frac{8}{3}D$ and $F^{\rm D} = 20F_0^{\rm D}$.
  The horizontal axis stands for the frequency normalized
  by the fundamental wave (from (a) to ($\text{a}^\prime$) in
  Fig.~\ref{fig:sche_p3}), $f_1$.
  The vertical axis stands for the intensity of each mode normalized
  by the peak intensity at $f=0$.
  The inset is the magnified {figure} in the vertical direction.
}
\end{figure}

\begin{figure}
  \includegraphics[width=\linewidth]{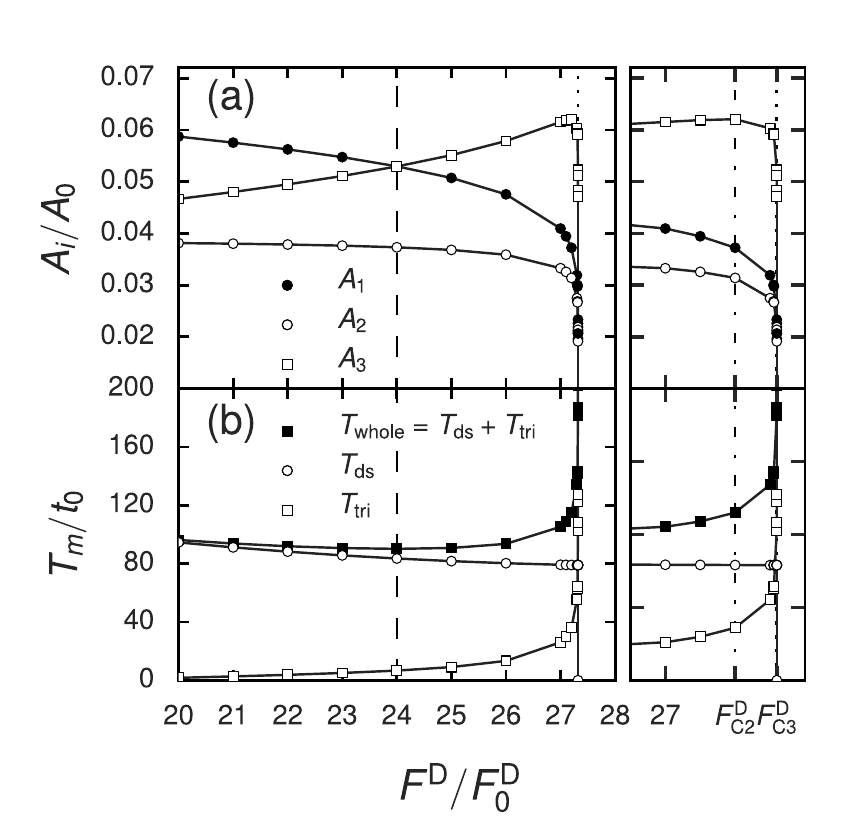}
  \caption{\label{fig:AC} Fourier spectrum of the
    time evolution of the limit cycle velocity in the doublet-singlet mode for $W=\frac{8}{3}D$ as functions of $F^{\rm D}$.
    (a) Amplitude of the first three essential Fourier modes ($f/f_1=1,2,3$) normalized by $A_0\equiv \tilde{v}\left( 0\right)$.
    (b) The whole cycle, $T_{\rm whole}$, the residence time
    in the triplet state, $T_{\rm tri}$, and
    that in the doublet-singlet state, 
    $T_{\rm ds}$, in the unit of $t_0$.
    The dashed lines stand for $F^{\rm D}_{\rm C1}=24F^{\rm D}_0$, the critical value of
      $F^{\rm D}$ where the crossover between $A_1$ and $A_3$ is observed
  and $T_{\rm whole}$ shows the minimum as well.
  The dotted line denotes $F^{\rm D}_{\rm C3}=27.32F^{\rm D}_0$, the critical value of
  the transition of the dynamical modes.
  The plots in the right panel show the magnified images at approximately $F^{\rm D}_{\rm C3}$
  in the $F^{\rm D}$-axis.
  The dash-dotted lines in the magnified plots denote $F^{\rm D}_{\rm C2}=27.2F^{\rm D}_0$ above which $A_3$ starts to decrease.
  }
\end{figure}

To clarify the behavior
around the dynamical mode transitions observed in our system,
we performed a spectrum analysis on the time evolution
of the velocity of a tagged particle
by using the following equation:
\begin{align}
  \tilde{v} \left( f \right) =  \int_{-\infty}^{\infty}e^{i2\pi f
    t}v\left( t\right)dt,\label{eq:fourier}
\end{align}
where $v\left( t\right)$ is the velocity of the tagged particle and $f$ is the frequency.
Because we confirmed that the altering of $W$
does not yield any essential change in the
qualitative behavior of the system, in the following,
we analyze the dynamics of the system with $W=8D/3$ as a typical case. 
In Fig.~\ref{fig:fourier}, we show the absolute value of the power
spectra obtained by
eq.(\ref{eq:fourier}) for $F^{\rm D}=20F_0^{\rm D}$.
The spectra of essential frequencies can be recognized as the
fundamental frequency $f_1$ and its harmonics.
Hereafter, we refer to the spectrum at $f=0$ as the zeroth spectrum,
and to the other essential spectra as
the first, second, third $\cdots$ in increasing order of frequency.
The zeroth spectrum reflects the average velocity.
The first essential spectrum comes from the whole cycle with three peaks,
while the third spectrum reflects the cycle between neighboring triplet
peaks, such as (b) and (d) in Fig.~\ref{fig:sche_p3}.
Because the characteristic three-peaked shape of the velocity time evolution can be almost
reproduced by using the three spectra of $f_1$, $f_2$ and $f_3$, these spectra can be regarded as the essential spectra.
  It is difficult to give a clear physical interpretation for each peak since the three-peaked pattern depends on all three spectra, but
  we can say that the third spectrum at least reflects
the difference in the magnitude of the spectrum between the average
velocity and the fastest velocity of the triplet cluster.
We show
the amplitudes of the first three
spectra, $A_i\equiv \left|\tilde v(i \cdot f_1)\right| \left( i \in
{1, 2, 3}\right)$, as functions of  $F^{\rm D}$ in Fig.~\ref{fig:AC}(a).
In Fig.~\ref{fig:AC}(b) we also show $T_{\rm tri}$,
the total residence time in the triplet
state during one whole cycle~((a)-($\text{a}^\prime$) in Fig.~\ref{fig:sche_p3}), and
$T_{\rm ds}$, the one in the doublet-singlet state.
The period of one whole cycle $T_{\rm whole}=T_{\rm ds} +T_{\rm tri}$ is also plotted.
In Fig.~\ref{fig:AC}(a), the amplitudes of $A_i$ are
normalized by $A_0\equiv \tilde v(0)$.
As $F^{\rm D}$ increases, $A_1$ decreases and $A_3$ increases.
Although $A_3$ is smaller than $A_1$ when $F^{\rm D}$ is smaller
than a threshold value $24F^{\rm D}_0$, above the threshold value, $A_3$ exceeds $A_1$.
That the third spectrum is greater than the first 
reflects the fact that the average velocity is higher than
the minimum value of the doublet velocity ((c) or
(e) in Fig.~\ref{fig:sche_p3}).
This crossover occurs because the
contribution of the triplet state becomes more dominant as
$T_{\rm tri}$ becomes larger.
Interestingly,
at around $F^{\rm D} = 24F^{\rm D}_0$, 
$T_{\rm whole}$ also exhibits a shallow minimum.
The non-monotonicity in $T_{\rm whole}$ can be understood by considering the
contributions
from $T_{\rm ds}$ and $T_{\rm tri}$ separately.
If the average velocity of the particles becomes faster,
the time needed by the
doublet to catch up to the singlet in front becomes shorter, which results in a
smaller $T_{\rm ds}$~(see Fig.~\ref{fig:AC}~(b)).
At the same time, 
as $F^{\rm D}$ becomes larger, the triplet cluster
becomes more stable, and $T_{\rm tri}$ becomes longer.
These two effects are in a trade-off relation, and the latter
dominates the system when $F^{\rm D}\ge 24F^{\rm D}_0$.
This value $F^{\rm D}_{\rm C1} = 24F^{\rm D}_0$ is referred to
as the first critical value, and the system has two more
critical values.
The second critical value is
$F^{\rm D}_{\rm C2} = 27.2F^{\rm D}_0$, where
$A_3$ shows a maximum, and above $F^{\rm D}_{\rm C2}$, the amplitude of the third spectrum, $A_3$, decreases abruptly.
The existence of this second critical force can be explained as follows.
As $F^{\rm D}$ approaches $F^{\rm D}_{\rm C2}$,
the residence time in the
triplet state becomes longer very sensitively to the change in $F^{\rm D}$,
although the velocity does not change greatly.
This leads to a dominant contribution of $A_3$
to the zeroth peak amplitude $A_0$ which reflects the average velocity.
In other words, the average velocity approaches the triplet velocity.
Because $A_3$ reflects the difference in the average velocity and the fastest velocity of the triplet state, the magnitude of the spectra decreases upon increasing $F^{\rm D}$ above
$F_{\rm C2}^{\rm D}$.
The third critical force is $F^{\rm D}_{\rm C3} = 27.32F^{\rm D}_0$, where the dynamical mode transition between the doublet-singlet mode and the triplet mode occurs.
As shown in Fig.~\ref{fig:AC}(b), 
the dynamical mode transition can be understood as a continuous
divergence of the residence time in the triplet state $T_{\rm tri}$
($T_{\rm ds}$ changes monotonously and almost linearly decreases with $F^{\rm D}-F^{\rm D}_{\rm C3}$).
Therefore, we characterize the divergence
in analogy with the second order phase transition
by using the following equation:
\begin{align}
  T_{\rm tri} = T_{\rm tri}^{\rm C}\left( \frac{F^{\rm D} - F^{\rm D}_{\rm C3}}{ F^{\rm D}_{\rm C3}}\right)^{-\beta},\label{eq:fit}
\end{align}
where
$T_{\rm tri}^{\rm C}$ and $\beta$
are constants and are found to be
  $T_{\rm tri}^{\rm C}\approx 17.9$,
$\beta \approx 0.15$ by a fitting of Eq.~(\ref{eq:fit}) using the least squares method to the residence time $T_{\rm tri}$.

\begin{figure}
  \begin{center}
\includegraphics[width=\linewidth]{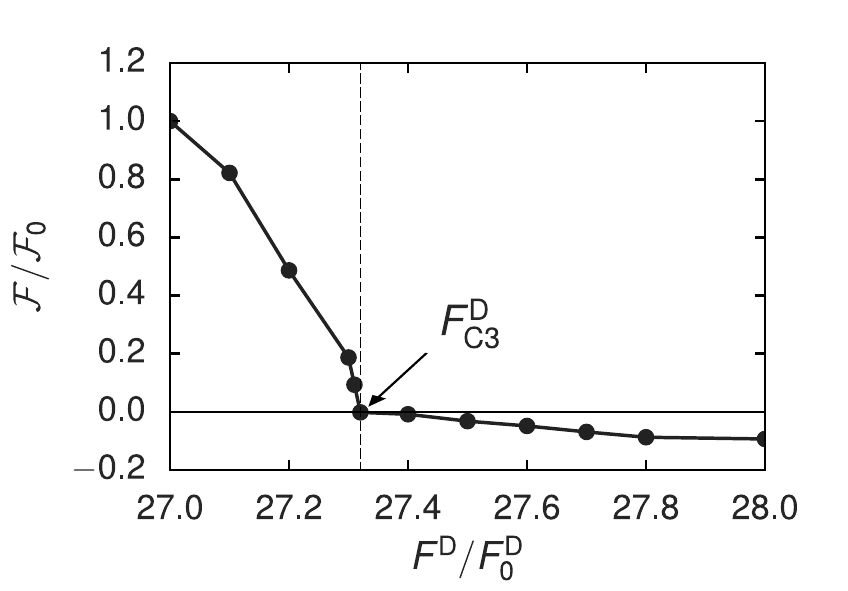}
\caption{\label{fig:EF}
  The evaluation function ${\cal F}$ as a function of $F^{\rm D}$.
  The values of ${\cal F}$ are normalized by ${\cal F}_0={\cal F}\left(27F_0^{\rm D}\right)$.
}
  \end{center}
\end{figure}

The transition of the dynamical modes can also be understood from the view
point of the stability of the triplet state.
When the triplet state breaks up 
and the doublet leaves the singlet behind,
the doublet has
a positive relative acceleration to the singlet.
Therefore, we consider the relative acceleration of the doublet to the singlet.
The equations of motion for three individual particles are as follows:
\begin{align}
  M_{\rm p} {\boldsymbol{a}_1} &= \bF^{\rm D}+\bF_1^{\rm H}+\bF_{12}^{\rm P},\label{eq:a1}\\
  M_{\rm p} {\boldsymbol{a}_2} &= \bF^{\rm D}+\bF_2^{\rm H}+\bF_{21}^{\rm P}+\bF_{23}^{\rm P},\label{eq:a2}\\
  M_{\rm p} {\boldsymbol{a}_3} &= \bF^{\rm D}+\bF_3^{\rm H}+\bF_{32}^{\rm P}\label{eq:a3},
\end{align}
where the particle indexes $i$=1,2,3 are assigned to the particles
from the front to the rear
in order,
$\boldsymbol{a}_i$ stands for the acceleration of particle $i$, and
$\bF_{ij}^{\rm P}$ represents the particle-particle direct interaction force exerted
on particle $i$ by particle $j$.
Now, we would like to consider the motion of the front doublet
(composed of particles 1 and 2) relative to
the trailing singlet (particle 3).
As the motion of the front doublet, we can simply consider the motion of
the center of mass of the two particles in the front (particles 1 and 2).
Since the action reaction relation $\bF_{12}^{\rm P}+\bF_{21}^{\rm P}=\boldsymbol{0}$ holds,
the summation of both sides of Eqs.~(\ref{eq:a1}) and (\ref{eq:a2})
leads to the following equation:
\begin{align}
  M_{\rm p}{\boldsymbol{a}_{\rm d}} &= \bF^{\rm D}+\frac{\bF_{23}^{\rm P}}{2}+\frac{\bF_1^{\rm H}+\bF_2^{\rm H}}{2}\label{eq:a4},
\end{align}
where $\boldsymbol{a}_{\rm d}=\left( \boldsymbol{a}_1+\boldsymbol{a}_2\right)/2$ stands for the acceleration of the doublet.
Then, the acceleration of the doublet relative to the singlet can be expressed by using Eqs.~(\ref{eq:a3}) and (\ref{eq:a4}), as follows:
\begin{align}
  M_{\rm p}\left({\boldsymbol{a}_{\rm d}-\boldsymbol{a}_{\rm s}} \right)&=
  \frac{\bF_{23}^{\rm P}}{2} - \bF_{32}^{\rm P} 
  +\frac{\bF_1^{\rm H}+\bF_2^{\rm H}}{2}
  -\bF_3^{\rm H},\label{eq:a5}
\end{align}
where we rename the acceleration of the singlet $\boldsymbol{a}_3$
as $\boldsymbol{a}_{\rm s}$.
As mentioned before, even when the triplet state is formed,
the three particles do not always have the exactly same velocity.
However, there is a time when the velocity of the leading doublet
(or the average velocity of the front two particles)
and that of the rear end particle coincide.
We will refer to this moment as the marginal point.
The stability of the triplet state can be quantified
by the relative acceleration at this marginal point.
For a measure for the relative acceleration,
now we introduce the following quantity at the marginal point:
\begin{align}
  {\cal F} = \left(
  \frac{\bF_{23}^{\rm P}}{2} - \bF_{32}^{\rm P} 
  +\frac{\bF_1^{\rm H}+\bF_2^{\rm H}}{2}
  -\bF_3^{\rm H}
  \right)\cdot
  \boldsymbol{e}_{\rm y}.
\end{align}
This is the $y$-component of the right-hand side of Eq.~(\ref{eq:a5}) at the marginal point.
In Fig.~\ref{fig:EF}, ${\cal F}$ is shown as a function of $F_{\rm D}$.
When ${\cal F}$ is positive,
the doublet has a positive acceleration relative to the singlet,
which leads to the breakage of the triplet state, {\it i. e.}, the doublet-singlet mode.
As ${\cal F}$ is positive but becomes decreases, 
the residence time in the triplet state becomes longer.
On the other hand,
when ${\cal F}$ is negative,
the singlet will further approach the doublet, and eventually, all three particles start
to move at the exactly same velocity, which results in the stable triplet state.
In agreement with the result of the spectrum analysis,
the change in the sign of ${\cal F}$ occurs at $F_{\rm C3}^{\rm D}=27.32F^{\rm D}$.

\begin{figure}
\includegraphics[width=\linewidth]{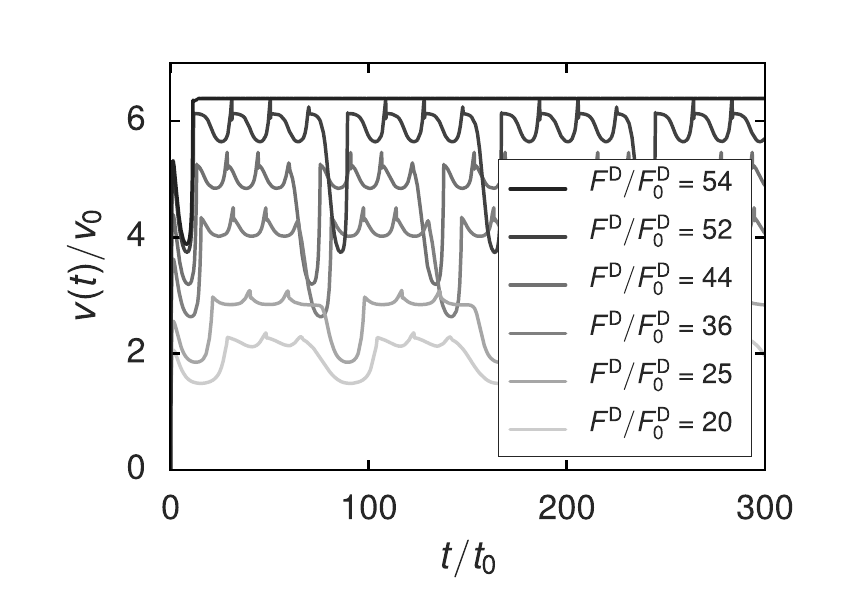}
\caption{\label{fig:p4_vel} Time evolution of the velocity of a tagged
  particle in the four-particle system with
  $W=\frac{8}{3}D$ for several $F^{\rm D}$s.}
\end{figure}

\begin{figure*}
  \begin{center}
\includegraphics[width=\textwidth]{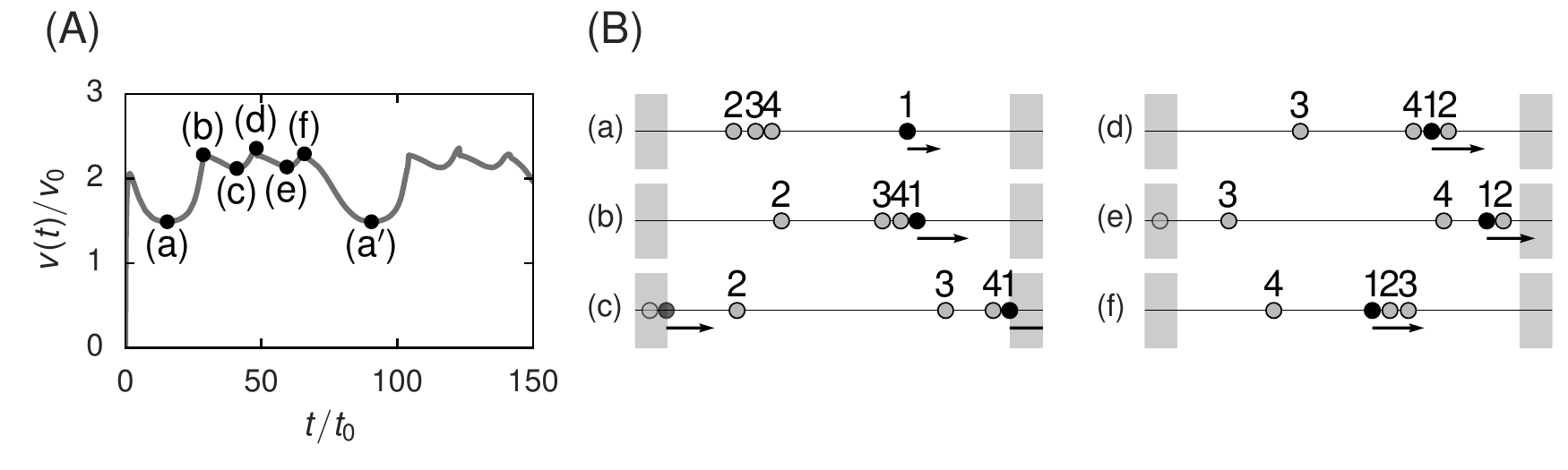}
\caption{\label{fig:sche_p4_211}
  (A) Time evolution of velocity of a tagged particle in four-particle system
  with $W=\frac{8}{3}D$ under $F^{\rm D}=20F^{\rm D}_0$,
  where the doublet-singlet-singlet mode appears.
  (B) Configuration of particles corresponding to (a)-(f) in (A). 
  All the elements of the pictures are the same as those in
  Fig.~\ref{fig:sche_p3}, but the scale of arrows is different.
  See also the supplementary movie S1.}
  \end{center}
\end{figure*}

\subsection{Four-particle system}

Similar to the three-particle system,
we performed simulations for a four-particle system for various values of $W$ and $F^{\rm D}$.
Again, the initial configuration is given to be a ``quartet'' in which
all the four particles gather in one cluster {with equal spacing and a small fluctuation between two adjacent particles.}
In Fig.~\ref{fig:p4_vel}, 
the time evolution of the velocity of a tagged particle is shown for 
$W=\frac{8}{3}D$ and several $F^{\rm D}$.
As shown in Fig.~\ref{fig:p4_vel}, we observe three distinct dynamical modes
in the four-particle system.
Similar to the three-particle system,
no qualitative difference can be seen in the dynamical behaviors even when the distance
between the walls is changed, as will be explained below.

\begin{figure*}
  \begin{center}
\includegraphics[width=\textwidth]{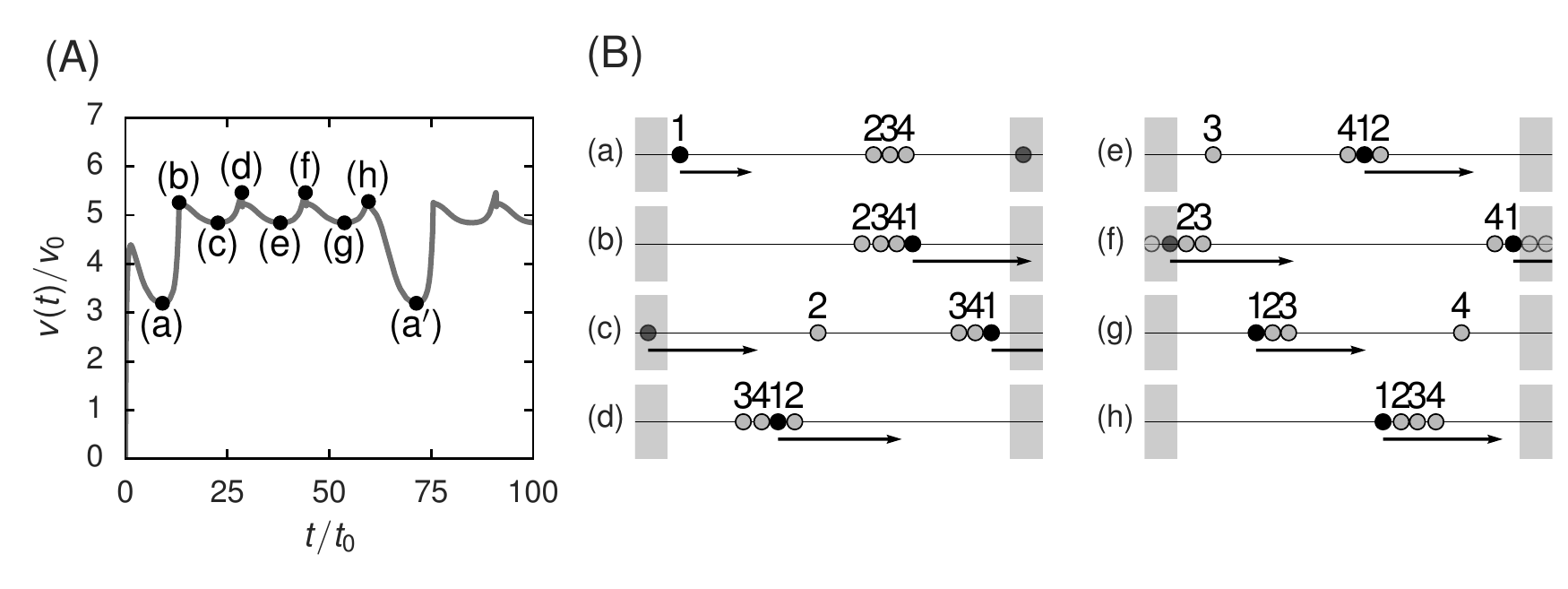}
\caption{\label{fig:sche_p4_31}
  (A) Time evolution of velocity of a tagged particle in a
  four-particle system
  with $W=\frac{8}{3}D$ under $F^{\rm D}=44F^{\rm D}_0$,
  where the triplet-singlet mode appears.
  (B) Configuration of particles corresponding to (a)-(h) in (A).
  All the elements are the same as those in Fig.~\ref{fig:sche_p4_211}.
  See also the supplementary movie S2.}
  \end{center}
\end{figure*}

When $F^{\rm D}\le 25F^{\rm D}_0$,
a mode with three peaks in one cycle appears.
The schematic snapshots
of the dynamical mode for $F^{\rm D}=20F_0^{\rm D}$ are shown in
Fig.~\ref{fig:sche_p4_211} (see also the supplementary movie S1).
The tops of the three peaks correspond to 
the triplet states ((b), (d) and (f)).
The triplet cluster is the fastest possible one in this mode
because the quartet cluster does not appear.
Similar to the doublet-singlet mode in the three-particle system,
when a triplet cluster is formed,
the leading two particles quickly detach,
resulting in the
doublet-singlet-singlet configuration ((c) and (e)).
Then, the doublet catches up to the trailing singlet, and the particles form a triplet again.
Changing the constituents of the doublet cluster, the system maintains
this cyclic motion.
The duration of the singlet state of the tagged particle as seen in (a)
is almost twice as long as that in the doublet
state ((c) and (e)).
This is because during the singlet state of the tagged particle (from (f) to (b) in the next cycle),
the other three particles form a triplet state that then breaks up into a doublet
and a singlet.
This process leads to three peaks in a single cycle.
We call this dynamical mode the ``doublet-singlet-singlet mode''.
Note that even though the initial configuration is a quartet state,
the system does not show any quartet state after the initial state.

When $F^{\rm D}$ is in the range of $25F^{\rm D}_0 <F^{\rm D}\le 52F^{\rm D}_0$, 
the time evolution of the velocity of a tagged particle 
shows four peaks in a single cycle.
The schematic snapshots for this mode are shown in
Fig.~\ref{fig:sche_p4_31}, where the case with $F^{\rm D}=44F^{\rm
  D}_0$ is chosen as a typical one for the mode (see also the supplementary movie S2).
In this case, the triplet cluster becomes more stable than the doublet,
and the doublet cluster does not appear.
The dynamical mode is very similar to the doublet-singlet mode in the
three-particle system.
The four peaks correspond to the quartet state, which is the fastest in
the mode ((b), (d), (f) and (h)).
The quartet spontaneously breaks up into a triplet and a singlet ((a), (c), (e) and (g))
because the triplet state is more stable in this
mode.
Because the triplet cluster is faster than the singlet,
owing to the periodic boundary condition
it catches up with the singlet, and again, the quartet cluster is formed.
The system continues this cyclic configurational change, similar to
the doublet-singlet mode in the three-particle system.
We refer to this mode as ``the triplet-singlet mode''.

In cases where $F^{\rm D}$ is larger than $52F^{\rm D}_0$,
the initial quartet no longer breaks.
In this mode, similar to the triplet mode in the three-particle system,
the particles form the quartet cluster, which moves stably with a constant velocity,
as shown in the case with $F^D = 54 F_0^D$ in Fig.~\ref{fig:p4_vel}.
The mode is referred to as ``the quartet mode''.

\begin{figure}
\includegraphics[width=\linewidth]{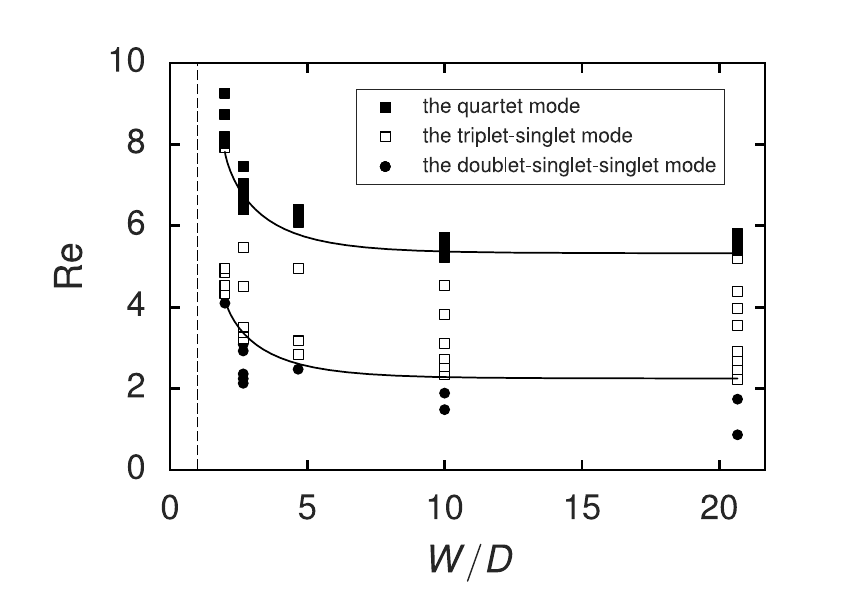}
\caption{\label{fig:p4_phase_diagram}
  Phase diagram for the
  four-particle system. Filled circles stand for
  doublet-singlet-singlet mode,
  blank squares represent the triplet-singlet mode and the filled squares represent the
  quartet mode. The solid lines are drawn
  as guides to show the boundaries of the dynamical modes.}
\end{figure}

To clarify the threshold Reynolds numbers at which
the transitions between different
dynamical modes occur,
we performed simulations of systems with
various $F^{\rm D}$ and $W$ values in the range  $2D \le W \le \frac{62}{3}D$.
The resultant phase diagram is shown in Fig.~\ref{fig:p4_phase_diagram}.
The features of the phase boundary lines are almost the same
as that obtained in the three-particle system:
the threshold Reynolds numbers of two transitions are constant for
$W \ge 10D$ and increase as $W$ decreases.
The phase boundary line between the doublet-singlet-singlet mode (where the
doublet state is the most stable) and the
triplet-singlet mode (where the triplet is the most stable)
is almost the same as that in
the three-particle system.
We refer to the boundary line between the triplet-singlet and the doublet-singlet-singlet modes
as the ``T-D line''.
The phase boundary line between   
the quartet mode and the triplet-singlet mode (``Q-T line'')
is located at a position with a Reynolds number that is nearly equidistant to the T-D transition Reynolds number for any $W$.
Although we did not perform the simulations for the systems with
more than four particles, one can expect that
the phase boundary lines, for example, the Quintet-Quartet (``Q-Q'') line in the five-particle system, show almost constant shifts from the T-D or Q-T lines with the same shape. 
Interestingly, the T-D line for a large enough $W$ again coincides with the threshold between the
Stokes regime and the Allen regime.

Lastly, we note that in the four-particle systems,
the collective behavior observed in the present work
is qualitatively different from that
reported in the experiment\cite{Sassa2012},
even at very low Reynolds numbers.
In the experiment on four-particle systems with a circular path\cite{Sassa2012},
the system exhibits the doublet-singlet-singlet mode, which is
observed in our systems, and after a while, the dynamical
mode changes into ``the doublet-doublet mode,'' where the particles
form two distinct doublet clusters and move stably.
In our calculations, neither the mode transition nor the
doublet-doublet mode was observed.
If the driving path is circular, the optical vortex itself and the resultant hydrodynamic interactions can exert torques that might play an important role in the collective behavior of particles.
It would be possible that the difference in the paths
between the experiment\cite{Sassa2012} and the present simulation leads to
different stable dynamical states and transitions between them.
To verify this hypothesis,
we also performed simulations with a circular path in which we applied a tangential driving force but not an external torque to particles.
Similar to the systems presented above,
the systems with a circular path are confined by two flat parallel walls,
and according to the experiment in Ref.~\cite{Sassa2012}, the ratio of the diameters of the
particle to that of the path is set to be $D/D_{\rm path}=3/10$, where $D_{\rm
  path}$ stands for the diameter of the path.
The results of our simulations showed
that the dynamical mode transition that was observed in the experiment
does not occur.
In the experiment, there are many possible factors that can bring different results, {\it e.g.},
the inhomogeneity in the light intensity,
particles not having exactly the same shape and size,
the effects of thermal fluctuations, the surface charge of the particles
and many other sources of complexity, which are not considered in our simulations.
Further investigation on the transitions between modes for comparison with experimental results
is beyond the scope of this paper, and we do not go into more
detail about this and leave these inquiries for future work.

\section{Conclusion}\label{sec:conc}
In the present paper,
we studied the collective dynamics of colloidal
particles immersed in a viscous fluid and driven by an external force along
a fixed straight trail by means of three dimensional direct numerical
simulations with fully resolved hydrodynamics.
The systems are confined by two flat parallel walls in the $z$ direction
and are periodic in the $x$ and $y$ directions in a Cartesian coordinate system.
Three- and four-particle systems are investigated, changing
the distance between the confining walls and the strength of the driving
force.
The results showed that the systems exhibit different dynamical modes depending on the Reynolds number and the distance between the walls.
The number of dynamical modes depends on the number of particles:
the three-particle system has two dynamical modes: the doublet-singlet mode and the triplet mode, while the
four-particle system has three modes: the doublet-singlet-singlet mode, the triplet-singlet mode and the quartet mode.
The critical value of the Reynolds number between different dynamical
modes becomes larger because of the increasing suppression of the hydrodynamic effect by the walls as the distance between walls becomes narrower.
However, this critical value approaches a constant value if the separation of walls is large enough to ignore the effects of the walls.
By using spectral decomposition of the time evolution of the velocity of a tagged particle, we analyzed the dynamical mode transition between the doublet-singlet mode and the triplet mode in three-particle system by assessing the diverging residence time in the triplet state in analogy with the second-order phase transition.
Such stability of the triplet state is also discussed from the view point of the total force acting on each particle.

The possible relation between the threshold between the Stokes regime
and the Allen regime in the single-particle system, ${\Re}\approx2$ and that
between the dynamical modes in our systems is intriguing.
In our system,
the transition reflects the change of the size of the most stable cluster,
while in a single-particle system, the transition in the behavior of the hydrodynamic drag force is due
to the appearance of nonlinearity.
We expect that our findings can be understood as the extension of the concept of the
Allen regime to multi-particle systems.
To give a definite conclusion, however, we have to clarify the underlying relation
between the nonlinearity in a single-particle system and the many-particle synchronization.
Such a relation remains an open question.

\section{Acknowledgment}
We would like to thank Yasuyuki Kimura
and Hiroaki Ito for enlightening discussions.
This work was supported by the Japan Society for the Promotion
of Science (JSPS) KAKENHI Grant No. 15H03708, No. 17H01083.


\begin{thebibliography}{32}
\bibitem[{not({\natexlab{a}})}]{note:rotation}
\bibinfo{note}{Particles are allowed to rotate freely, but in the present
  situations, no torque is exerted on particles because of the symmetry of the
  system}.
\end{thebibliography}

\end{document}